\documentclass[aps,prl,twocolumn,superscriptaddress,showpacs,preprintnumbers]{revtex4-1}
\usepackage{lmodern}
\usepackage{amsmath,amssymb,amsthm} 
	
\usepackage{grffile} 
\usepackage{sidecap} 
\usepackage{enumitem} 
\usepackage{array}
\usepackage{multirow}
\usepackage{ctable} 
\usepackage{subfigure}	
\usepackage{hyperref}
\usepackage{lineno}
\usepackage{etoolbox}
\usepackage{textcomp}
\usepackage{indentfirst}
\newcommand{\GeV}{$\mathrm{GeV}/{c^{2}}$}
\newcommand{\MeV}{$\mathrm{MeV}/{c^{2}}$}

\newcommand{\PPRHC}{$B^+ \rightarrow p\overline{p}\rho^+ $}
\newcommand{\PPPIPI}{$B^0 \rightarrow p\overline{p}\pi^+\pi^- $}
\newcommand{\PPPIPIC}{$B^+ \rightarrow p\overline{p}\pi^+\pi^0 $}

\newcommand{\PPK}{$B^0 \rightarrow p\overline{p}K^0 $}

\begin{document}


\title{Study of \boldmath{$B \rightarrow p\overline{p}\pi\pi$}}

\noaffiliation
\affiliation{University of the Basque Country UPV/EHU, 48080 Bilbao}
\affiliation{Brookhaven National Laboratory, Upton, New York 11973}
\affiliation{Budker Institute of Nuclear Physics SB RAS, Novosibirsk 630090}
\affiliation{Faculty of Mathematics and Physics, Charles University, 121 16 Prague}
\affiliation{Chonnam National University, Gwangju 61186}
\affiliation{University of Cincinnati, Cincinnati, Ohio 45221}
\affiliation{Deutsches Elektronen--Synchrotron, 22607 Hamburg}
\affiliation{Duke University, Durham, North Carolina 27708}
\affiliation{University of Florida, Gainesville, Florida 32611}
\affiliation{Key Laboratory of Nuclear Physics and Ion-beam Application (MOE) and Institute of Modern Physics, Fudan University, Shanghai 200443}
\affiliation{II. Physikalisches Institut, Georg-August-Universit\"at G\"ottingen, 37073 G\"ottingen}
\affiliation{SOKENDAI (The Graduate University for Advanced Studies), Hayama 240-0193}
\affiliation{Gyeongsang National University, Jinju 52828}
\affiliation{Department of Physics and Institute of Natural Sciences, Hanyang University, Seoul 04763}
\affiliation{University of Hawaii, Honolulu, Hawaii 96822}
\affiliation{High Energy Accelerator Research Organization (KEK), Tsukuba 305-0801}
\affiliation{J-PARC Branch, KEK Theory Center, High Energy Accelerator Research Organization (KEK), Tsukuba 305-0801}
\affiliation{Forschungszentrum J\"{u}lich, 52425 J\"{u}lich}
\affiliation{IKERBASQUE, Basque Foundation for Science, 48013 Bilbao}
\affiliation{Indian Institute of Science Education and Research Mohali, SAS Nagar, 140306}
\affiliation{Indian Institute of Technology Bhubaneswar, Satya Nagar 751007}
\affiliation{Indian Institute of Technology Guwahati, Assam 781039}
\affiliation{Indian Institute of Technology Hyderabad, Telangana 502285}
\affiliation{Indian Institute of Technology Madras, Chennai 600036}
\affiliation{Indiana University, Bloomington, Indiana 47408}
\affiliation{Institute of High Energy Physics, Chinese Academy of Sciences, Beijing 100049}
\affiliation{Institute of High Energy Physics, Vienna 1050}
\affiliation{INFN - Sezione di Napoli, 80126 Napoli}
\affiliation{INFN - Sezione di Torino, 10125 Torino}
\affiliation{Advanced Science Research Center, Japan Atomic Energy Agency, Naka 319-1195}
\affiliation{J. Stefan Institute, 1000 Ljubljana}
\affiliation{Institut f\"ur Experimentelle Teilchenphysik, Karlsruher Institut f\"ur Technologie, 76131 Karlsruhe}
\affiliation{Kennesaw State University, Kennesaw, Georgia 30144}
\affiliation{King Abdulaziz City for Science and Technology, Riyadh 11442}
\affiliation{Department of Physics, Faculty of Science, King Abdulaziz University, Jeddah 21589}
\affiliation{Kitasato University, Sagamihara 252-0373}
\affiliation{Korea Institute of Science and Technology Information, Daejeon 34141}
\affiliation{Korea University, Seoul 02841}
\affiliation{Kyungpook National University, Daegu 41566}
\affiliation{LAL, Univ. Paris-Sud, CNRS/IN2P3, Universit\'{e} Paris-Saclay, Orsay 91898}
\affiliation{\'Ecole Polytechnique F\'ed\'erale de Lausanne (EPFL), Lausanne 1015}
\affiliation{P.N. Lebedev Physical Institute of the Russian Academy of Sciences, Moscow 119991}
\affiliation{Faculty of Mathematics and Physics, University of Ljubljana, 1000 Ljubljana}
\affiliation{Ludwig Maximilians University, 80539 Munich}
\affiliation{Luther College, Decorah, Iowa 52101}
\affiliation{University of Maribor, 2000 Maribor}
\affiliation{Max-Planck-Institut f\"ur Physik, 80805 M\"unchen}
\affiliation{School of Physics, University of Melbourne, Victoria 3010}
\affiliation{University of Mississippi, University, Mississippi 38677}
\affiliation{Moscow Physical Engineering Institute, Moscow 115409}
\affiliation{Moscow Institute of Physics and Technology, Moscow Region 141700}
\affiliation{Graduate School of Science, Nagoya University, Nagoya 464-8602}
\affiliation{Kobayashi-Maskawa Institute, Nagoya University, Nagoya 464-8602}
\affiliation{Universit\`{a} di Napoli Federico II, 80055 Napoli}
\affiliation{Nara Women's University, Nara 630-8506}
\affiliation{National Central University, Chung-li 32054}
\affiliation{National United University, Miao Li 36003}
\affiliation{Department of Physics, National Taiwan University, Taipei 10617}
\affiliation{H. Niewodniczanski Institute of Nuclear Physics, Krakow 31-342}
\affiliation{Nippon Dental University, Niigata 951-8580}
\affiliation{Niigata University, Niigata 950-2181}
\affiliation{Novosibirsk State University, Novosibirsk 630090}
\affiliation{Osaka City University, Osaka 558-8585}
\affiliation{Pacific Northwest National Laboratory, Richland, Washington 99352}
\affiliation{Panjab University, Chandigarh 160014}
\affiliation{University of Pittsburgh, Pittsburgh, Pennsylvania 15260}
\affiliation{Research Center for Nuclear Physics, Osaka University, Osaka 567-0047}
\affiliation{Theoretical Research Division, Nishina Center, RIKEN, Saitama 351-0198}
\affiliation{University of Science and Technology of China, Hefei 230026}
\affiliation{Showa Pharmaceutical University, Tokyo 194-8543}
\affiliation{Soongsil University, Seoul 06978}
\affiliation{Sungkyunkwan University, Suwon 16419}
\affiliation{School of Physics, University of Sydney, New South Wales 2006}
\affiliation{Department of Physics, Faculty of Science, University of Tabuk, Tabuk 71451}
\affiliation{Tata Institute of Fundamental Research, Mumbai 400005}
\affiliation{Department of Physics, Technische Universit\"at M\"unchen, 85748 Garching}
\affiliation{Earthquake Research Institute, University of Tokyo, Tokyo 113-0032}
\affiliation{Department of Physics, University of Tokyo, Tokyo 113-0033}
\affiliation{Tokyo Metropolitan University, Tokyo 192-0397}
\affiliation{Virginia Polytechnic Institute and State University, Blacksburg, Virginia 24061}
\affiliation{Wayne State University, Detroit, Michigan 48202}
\affiliation{Yamagata University, Yamagata 990-8560}
\affiliation{Yonsei University, Seoul 03722}
  \author{K.~Chu}\affiliation{Department of Physics, National Taiwan University, Taipei 10617} 
  \author{M.-Z.~Wang}\affiliation{Department of Physics, National Taiwan University, Taipei 10617} 
  \author{I.~Adachi}\affiliation{High Energy Accelerator Research Organization (KEK), Tsukuba 305-0801}\affiliation{SOKENDAI (The Graduate University for Advanced Studies), Hayama 240-0193} 
  \author{H.~Aihara}\affiliation{Department of Physics, University of Tokyo, Tokyo 113-0033} 
  \author{S.~Al~Said}\affiliation{Department of Physics, Faculty of Science, University of Tabuk, Tabuk 71451}\affiliation{Department of Physics, Faculty of Science, King Abdulaziz University, Jeddah 21589} 
  \author{D.~M.~Asner}\affiliation{Brookhaven National Laboratory, Upton, New York 11973} 
  \author{V.~Aulchenko}\affiliation{Budker Institute of Nuclear Physics SB RAS, Novosibirsk 630090}\affiliation{Novosibirsk State University, Novosibirsk 630090} 
  \author{T.~Aushev}\affiliation{Moscow Institute of Physics and Technology, Moscow Region 141700} 
  \author{R.~Ayad}\affiliation{Department of Physics, Faculty of Science, University of Tabuk, Tabuk 71451} 
  \author{V.~Babu}\affiliation{Deutsches Elektronen--Synchrotron, 22607 Hamburg} 
% ...
\collaboration{The Belle Collaboration}

\begin{abstract}
Using a data sample of $772 \times 10^6  B\overline B$
pairs collected on the $\Upsilon(4S)$ resonance with the Belle detector at the
KEKB asymmetric-energy $e^+ e^-$ collider,
we report the observations of \PPPIPI~and \PPPIPIC.  We measure a decay branching
fraction of $(0.83\pm0.17\pm0.17)\times 10^{-6}$ in \PPPIPI~for $M_{\pi^+\pi^-}<1.22$ \GeV~with a significance of 5.5 standard deviations.  The contribution from \PPK~is excluded.
We measure a decay branching fraction of $(4.58\pm1.17\pm0.67)\times 10^{-6}$
for \PPPIPIC~with $M_{\pi^+\pi^0}<1.3$ \GeV~with a significance of 5.4 standard deviations.
We study the difference of the $M_{p\overline{p}}$ distributions in \PPPIPI~and \PPPIPIC.
\end{abstract}


\maketitle

Charmless $B$ decays offer a good opportunity to find sizable CP violation due to interference between
the $b \rightarrow s$ penguin and $b \rightarrow u$ tree processes. Such decays can reveal new physics if measured results deviate
from Standard Model expectations. In the B-factory era, both Belle and BaBar have discovered large direct CP violation in
the $B \rightarrow K\pi$ system~\cite{REF36,REF35,Nature}.  The $\rm{LHCb}$ collaboration reported evidence of direct CP violation
in $B^+ \rightarrow p \overline{p} K^+$~\cite{REF31}.
Here and throughout the text, the inclusion of the charge-conjugate mode is implied unless otherwise stated.
This rare baryonic $B$ decay presumably proceeds via the $b \rightarrow s$ penguin process
with some non-negligible $b \rightarrow u$ contribution.
It is intriguing that the invariant mass of the $p\overline{p}$ system peaks
near threshold~\cite{REF40} and in the $p\overline{p}$ rest frame, $K^+$ is produced preferably
in the $\overline{p}$ direction.~\cite{REF24}.  Interestingly, this angular asymmetry is opposite to that
observed in $B^+ \rightarrow p\overline{p}\pi^+$ which is presumably dominated by the
$b \rightarrow u$ tree process~\cite{REF24}.
Most of the baryonic $B$ decays presumably proceed predominantly via the $b\rightarrow s$ process
except for $B^+\rightarrow p\overline{p}\pi^+$ and $B^0\rightarrow p\overline{p}\pi^0$~\cite{REF34} decays.
It is important to measure other $b \rightarrow u$ baryonic $B$ decays to provide more
information for theoretical investigation based on a generalized factorization
approach~\cite{REF29}.

We report a study of both \PPPIPI~and \PPPIPIC~including
the $B \rightarrow p\overline{p}\rho$ mass region using the full $\Upsilon(4S)$ data set collected by the Belle detector~\cite{Abashian02,Brodzicka12} at the asymmetric-energy $e^+$ (3.5 $\rm{GeV}$) $e^-$ (8 $\rm{GeV}$) KEKB collider~\cite{Kurokawa03,Abe13}. The data sample used in this study corresponds to an integrated luminosity of 711 fb$^{-1}$,
which contains $772\times10^{6}$ $B\overline{B}$ pairs produced on the $\Upsilon(4S)$ resonance.
The Belle detector surrounds the interaction point of KEKB. It is a large-solid-angle magnetic
spectrometer that consists of a silicon vertex detector, a 50-layer central drift chamber (CDC), an array of
aerogel threshold Cherenkov counters (ACC),  
a barrel-like arrangement of time-of-flight scintillation counters (TOF), and an electromagnetic calorimeter (ECL)
comprised of CsI(Tl) crystals located inside a superconducting solenoid coil that provides a 1.5~T
magnetic field.  An iron flux-return located outside of the coil is instrumented to detect $K_L^0$ mesons and identify
muons.

For the study of $B \rightarrow p\overline{p}\pi\pi$, samples simulated with the Monte Carlo technique (MC)
are used to optimize the signal selection
criteria and estimate the signal reconstruction efficiency. These samples are
generated with EvtGen~\cite{evtgen}, and a Geant~\cite{geant}-based software package to model
the detector response.  We generate the signal MC sample by a phase space model reweighted
with the $p\overline{p}$ mass distribution obtained by $\rm{LHCb}$~\cite{REF38}. The background samples include
the continuum events ($e^+e^-\rightarrow u\overline{u}$, $d\overline{d}$, $s\overline{s}$, and $c\overline{c}$), generic
$B$ decays ($b \rightarrow c$) and rare $B$ decays ($b \rightarrow u, d, s$). These simulated background
samples are six times larger than the integrated luminosity of the accumulated Belle data.

We require charged particles to originate within a $2.0$ cm region along the beam and from a $0.3$ cm region on the transverse plane around the interaction region. To identify charged particles, we utilize the
likelihood information determined for each particle type by the CDC, TOF and ACC and apply the same selection criteria
listed in~\cite{REF24} to select $p (\overline{p})$ and $\pi^+ (\pi^-)$. The $\pi^0$ is reconstructed
from two photons with a minimum energy in the laboratory frame of 0.05 $\rm{GeV}$ measured by the ECL.
To reduce combinatoric background, the $\pi^0$ energy is required to be greater than 0.5 $\rm{GeV}$
and the reconstructed mass is in the range 0.111$<$$M_{\gamma\gamma}$$<$0.151 \GeV, which corresponds to about a $\pm3.0$
standard deviation ($\sigma$) window. We then perform a mass-constrained fit to the nominal
$\pi^0$ mass~\cite{pdg18} in order to improve the resolution of the reconstructed $\pi^0$ four-momentum.
To reject $B\rightarrow p\overline{p}D^{(*)}$ events, we restrict the
invariant $\pi\pi$ mass $M_{\pi\pi}$ to be less than 1.22 \GeV~for \PPPIPI~and 1.3 \GeV~for \PPPIPIC~based on studies of the
simulated background.
We use $\Delta E=E^*_{\rm recon} - E^*_{\rm beam}$ and $M_{\rm bc}=\sqrt{(E^*_{\rm beam}/c^2)^2-(P^*_{\rm recon}/c)^2}$,
to identify $B$ decays.  $E^*_{\rm recon}$/$P^*_{\rm recon}$ and $E^*_{\rm beam}$ are the reconstructed $B$ energy/momentum
and the beam energy measured in the $\Upsilon (4S)$ rest frame, respectively.
For further investigation, we keep candidates with 5.24 $< M_{\rm bc} <$ 5.29 \GeV~and $|\Delta E| <$ 0.2 $\rm{GeV}$.

We have further applied a $D$ veto to reject candidate events with a charged pion, assumed to be a charged kaon,
satisfying $|M_{K\pi} – M_D| <$ 0.4 \GeV.
We require only one $B$ candidate in each event.
We choose the candidate with the smallest value of $\chi^2$ in the $B$ vertex fit. The fractions
of \PPPIPI~and \PPPIPIC~MC events with multiple $B$ candidates are 16.4\% and 20.3\%, respectively.
This selection removes  5.6\% of \PPPIPI~and 8.7\% of \PPPIPIC~signal.

Based on the MC simulation, there are only a few events from generic or rare $B$ decays in the candidate
region (5.27 $< M_{\rm bc} <$ 5.29 \GeV~and $|\Delta E| <$ 0.2 $\rm{GeV}$), thus they are ignored.
The continuum background is the dominant component in the candidate region.
Variables describing event topology are used to distinguish spherical $B \overline{B}$ events
from jet-like continuum events.  We use a neural network package, NeuroBayes~\cite{NB},
to separate the $B$ signal from the continuum background. There are 28 input parameters for
the neural network training, of which 23 parameters are modified Fox-Wolfram moments of
particles of the signal $B$ candidate, and separately those of particles in the rest
of the event~\cite{KSFW, KSFW2}.
The remaining five parameters are the separation between the $B$
candidate vertex and the accompanying $B$ vertex along the longitudinal direction; the angle between the $B$ flight
direction and the beam axis in the $\Upsilon (4S)$ rest frame;
the angle between $B$ momentum and the thrust axis of the event in
the $\Upsilon (4S)$ rest frame; the sphericity~\cite{Sper} of the event
calculated in the $\Upsilon (4S)$ rest frame; and the $B$ flavor tagging quality parameter~\cite{qr}.

\begin{figure}[htbp]
\footnotesize
\centering
{
\includegraphics[width=6cm]{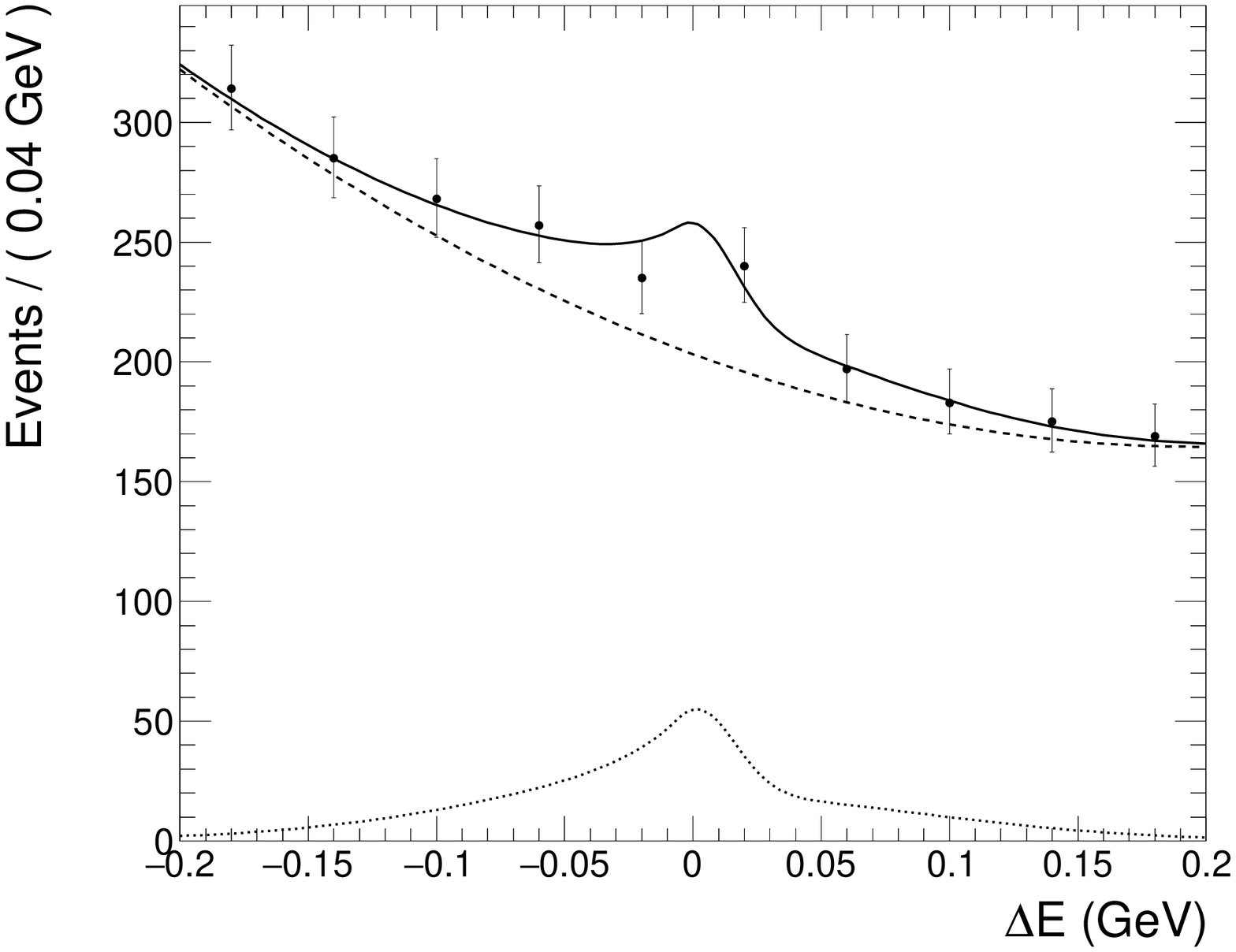}
\includegraphics[width=6cm]{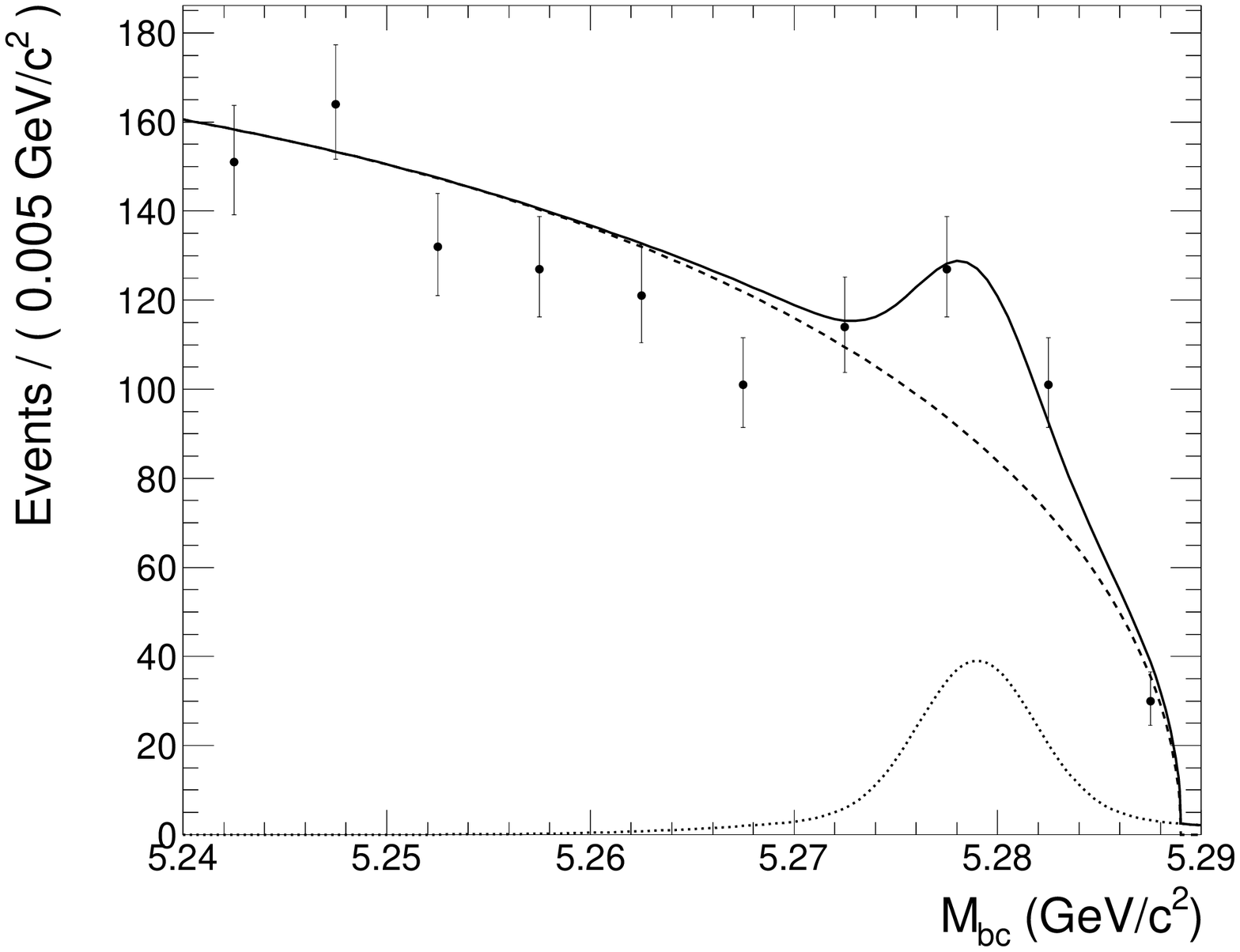}
}
\caption{Fit results of \PPPIPIC~projected onto $\Delta E$ (with 5.27 $< M_{\rm bc} <$ 5.29 \GeV) and
$M_{\rm bc}$ (with $-$0.03 $<\Delta E<$ 0.03 $\rm{GeV}$).
The dashed line represents the background. The dotted line represents the signal. The solid line is the sum of all fit components.}
\label{fig:ppr2fitting}
\end{figure}

The output of NeuroBayes, $C_{\rm nb}$, ranges from $-1$ to  $+1$, where the value is close to  $+1$
for $B\overline{B}$-like and  $-1$ for continuum-like events.
We require the $C_{\rm nb}$ to be greater than 0.9 (0.87) for \PPPIPI (\PPPIPIC) with optimizations based on
a figure-of-merit (FOM) defined as:
\begin{equation}
  \rm{FOM} = \frac{\textit{N}_s}{\sqrt{\textit{N}_s+\textit{N}_b}},
\end{equation}
where $\textit{N}_{\rm s}$ is the expected signal yield assuming the branching fraction measured by LHCb
for \PPPIPI, the same value for \PPPIPIC, and $\textit{N}_{\rm b}$ is the number of background events
from the MC simulations.
To extract the $B\rightarrow p\overline{p}\pi\pi$ yield for events in the candidate region, we perform an extended unbinned
likelihood fit to variables
$\Delta E$ and $M_{\rm bc}$.
These variables are assumed to be uncorrelated. The fit function used is:
\begin{equation}
\label{liklihood}
\mathcal{L}=\frac{e^{-\Sigma_{j=1}^2(N_{j})}}{N!}\prod_{i=1}^{N}\sum_{j}(N_{j} P_{j}(M_{\rm bc}^i, \Delta{E}^i)),
\end{equation}
where $N$ is the number of total events, $i$ denotes the event index, $j$ stands for the component index (signal or background), and $P$ represents
the probability density function (PDF).\\
\begin{figure}[htbp]
\footnotesize
\centering
{
\includegraphics[height=5cm,width=6cm]{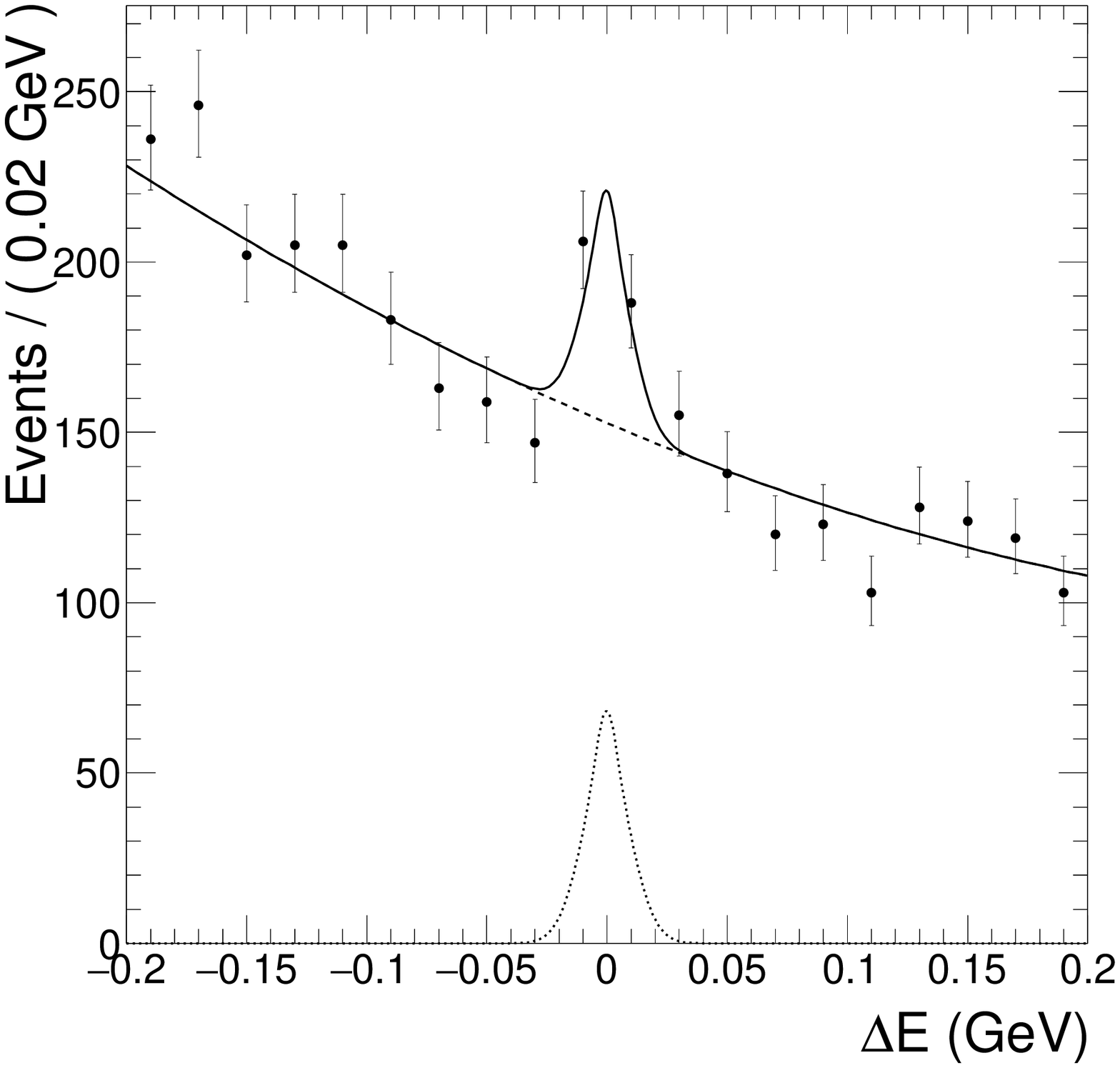}
\includegraphics[height=5cm,width=6cm]{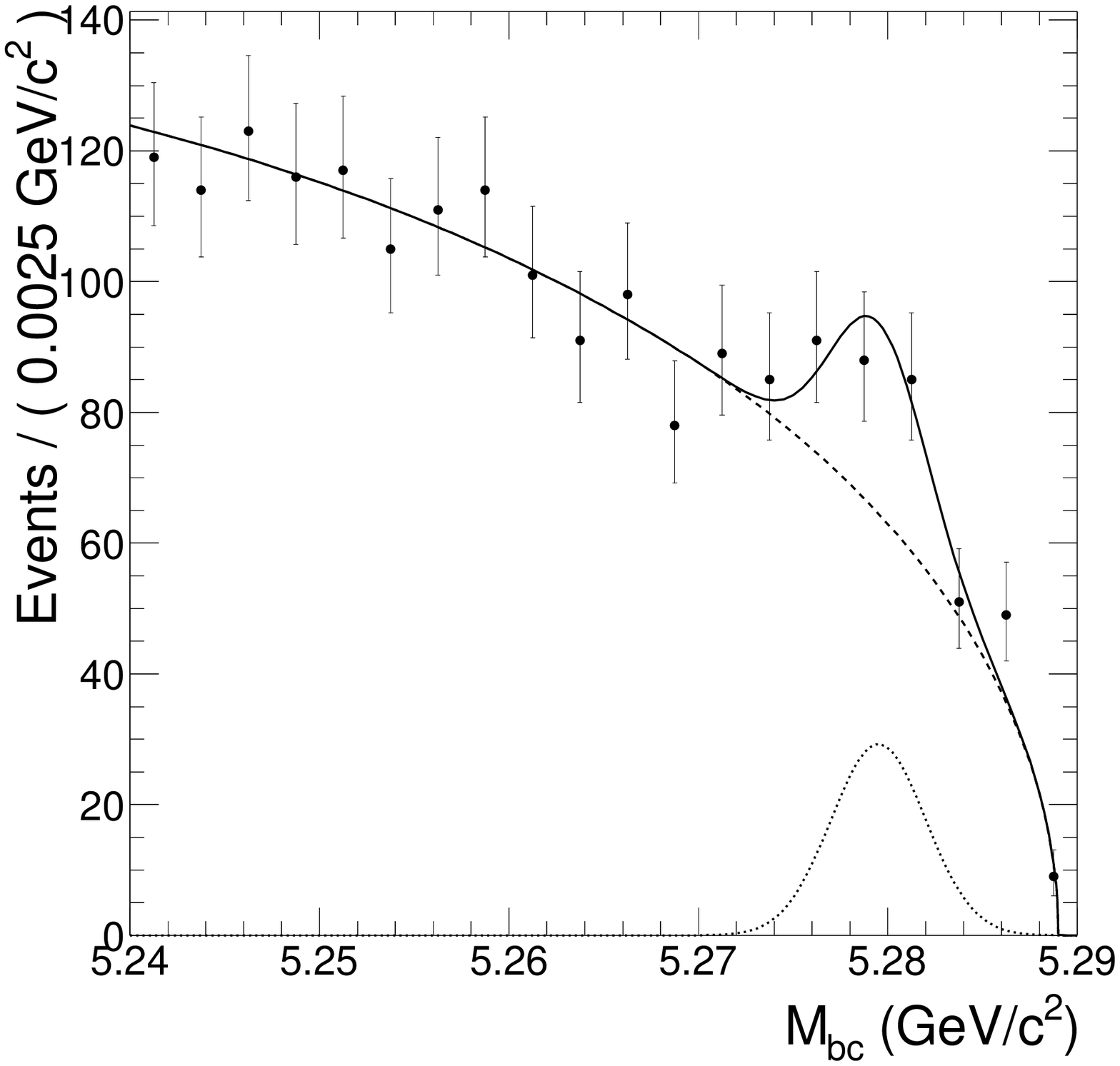}
}
\caption{Fit results of \PPPIPI~projected onto $\Delta E$ (with 5.27 $< M_{\rm bc} <$ 5.29 \GeV) and $M_{\rm bc}$ (with $-$0.03 $<\Delta E<$ 0.03 $\rm{GeV}$)
The dashed line represents the background. The dotted line represents the signal. The solid line is the sum of all fit components.}
\label{fig:pprhofitting}
\end{figure}
To model the signal distributions, we use a double Gaussian functions for $\Delta E$ of \PPPIPI, a Crystal Ball function~\cite{Crys} and a Gaussian function for $\Delta E$ of \PPPIPIC, and a double Gaussian function for $M_{\rm bc}$. For the background, we use a second-order Chebyshev polynomial function and an ARGUS function~\cite{argus}
to describe $\Delta E$ and $M_{\rm bc}$, respectively.
The signal distributions in $\Delta E$ and $M_{\rm bc}$ are calibrated with
the $B^0 \rightarrow p\overline{p}\overline{D}{}^0$ ($\overline{D}{}^0\rightarrow K^+\pi^-$)
and $B^0 \rightarrow \overline{D}{}^0 \pi^0$ ($\overline{D}{}^0\rightarrow K^+\pi^-$) by comparing
the shape difference between the prediction of the MC and data.  These modes have the same multiplicity in the
final state as our signal, much larger statistics, and small backgrounds.
We fix the calibrated signal shapes from MC simulation and allow the component yields
and all other PDF shape parameters to float. The fit results are shown in Figs.~\ref{fig:pprhofitting}
and~\ref{fig:ppr2fitting}.

\begin{figure}[htbp]
\footnotesize
\centering
{
\includegraphics[width=6cm]{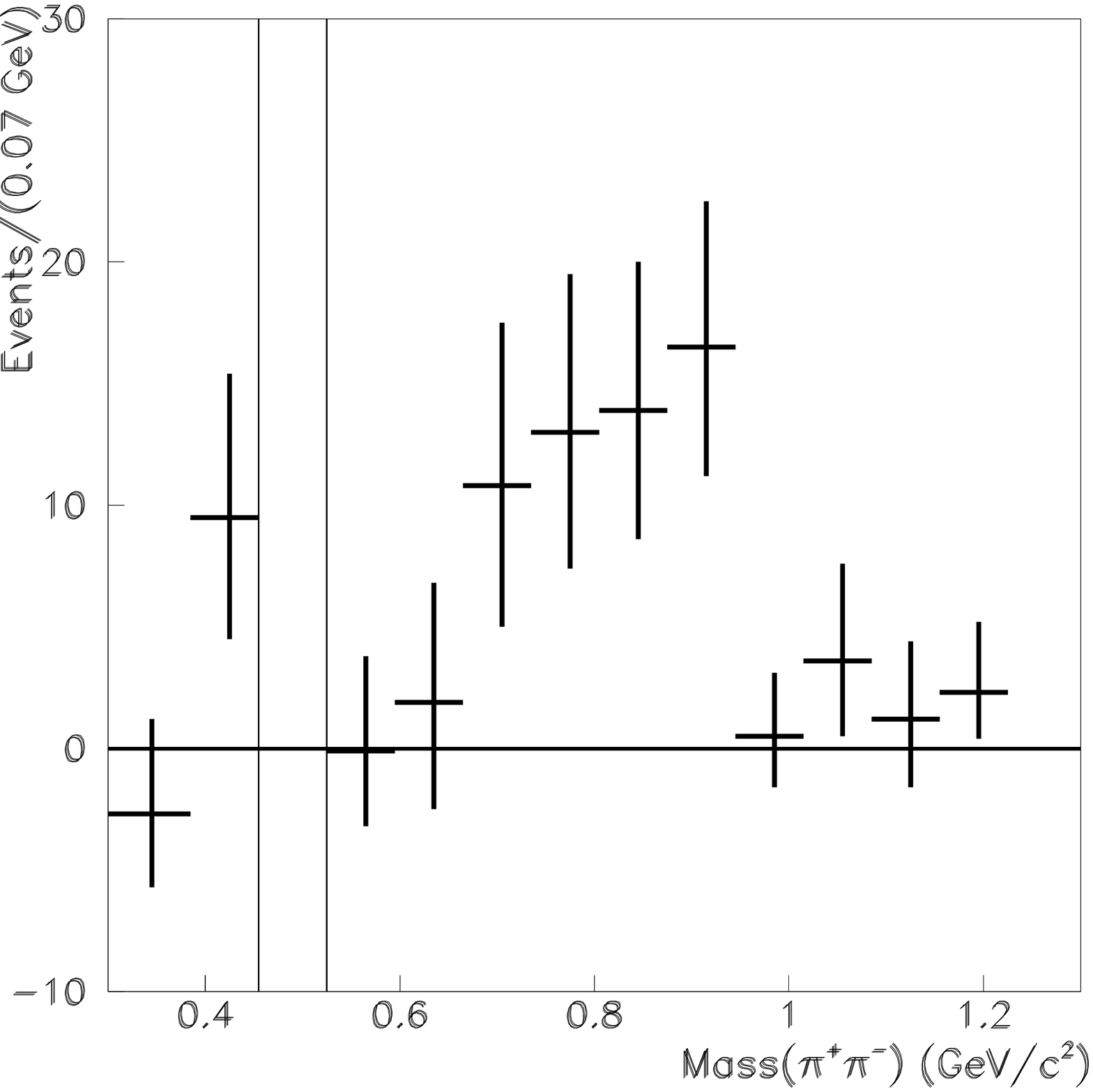}}

\caption{The $M_{\pi\pi}$ signal distribution for \PPPIPI.}
\label{fig:impipi}
\end{figure}

\begin{figure}[htbp]
\footnotesize
\centering
{
\includegraphics[width=6cm]{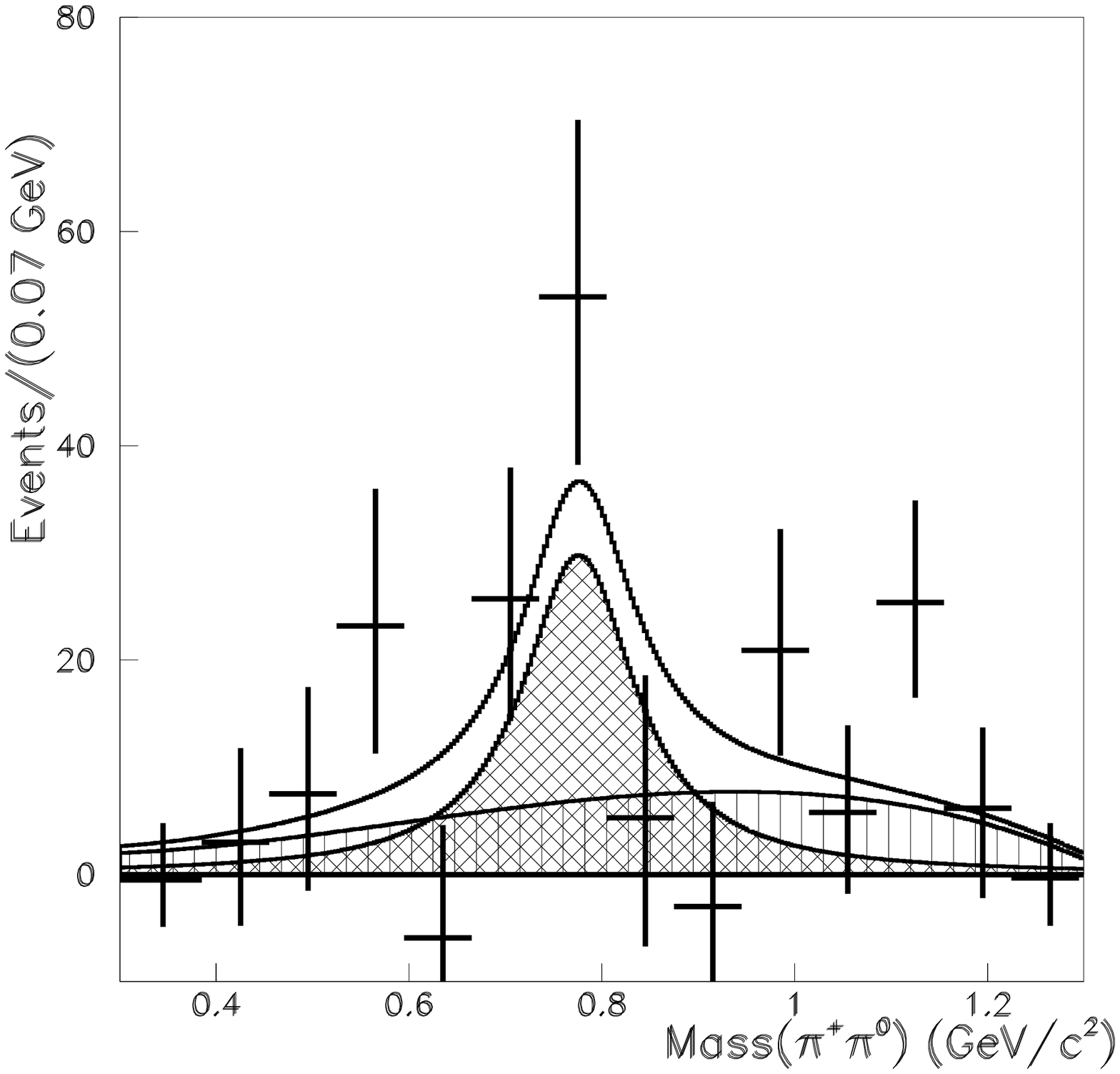}}
\caption{Fit results of \PPPIPIC~in different $M_{\pi\pi}$ bins, the cross hatched region represents \PPRHC~component and the vertical line hatched region represents \PPPIPIC~component.}
\label{fig:ppr2chi}
\end{figure}

We find signal yields of \PPPIPI~and \PPPIPIC~to be $73.8^{+15.8}_{-14.9}$ and $151\pm39$ with a fit significance of 5.5$\sigma$ and 5.4$\sigma$, respectively.
 The significance is defined as
$\sqrt{-2\times\ln({\mathcal{L}_0}/{\mathcal{L}_s})} (\sigma)$,
where $\mathcal{L}_0$ is the likelihood with zero signal yield and $\mathcal{L}_s$ is the likelihood for the measured yield.
In this calculation, we have used the likelihood function which is smeared by including
the additive systematic uncertainties that affect the yield.
With the large significance of both modes we then measure the signal
yields in different $M_{\pi\pi}$ bins with the same fit method.
Table~\ref{table:impipipprho} and Fig.~\ref{fig:impipi} show the yield and statistical significance in
different $M_{\pi\pi}$ bins for \PPPIPI and Table~\ref{table:impipippr2}/Fig.~\ref{fig:ppr2chi} for \PPPIPIC.
For \PPPIPI, signal events in the bin $0.46<M_{\pi\pi}<0.53$ \GeV~are mostly
from $B^0\rightarrow p\overline{p}K_S^0$, and hence we exclude this range in the contribution shown in
Table~\ref{table:impipipprho} and Fig.~\ref{fig:impipi} and from the measurement of $\mathcal{B}$(\PPPIPI).
Assuming the $\Upsilon(4S)$ decays to charged and neutral $B\bar{B}$ pairs equally,
we use the efficiency obtained from the MC simulation and fitted signal yield to calculate the branching fraction.
After calculating overall efficiencies for \PPPIPI~and \PPPIPIC, the
branching fractions of \PPPIPI~and \PPPIPIC~for $M_{\pi^+\pi^-}<1.22$ \GeV~and
$M_{\pi^+\pi^0}<1.3$ \GeV~are found to be $(0.83\pm0.17\pm0.17)\times 10^{-6}$ and $(4.58\pm1.17\pm0.67)\times 10^{-6}$; the signal efficiencies are 11.5\% and 4.3\%, respectively.\\
\begin{table}[htbp]
\begin{center}
\caption{Yields, statistical significance and efficiencies ($\varepsilon_{\rm eff}$) in different $M_{\pi\pi}$ bin for \PPPIPI.}
\label{table:impipipprho}
\begin{tabular}[t]{lccc}
\hline\hline
$M_{\pi\pi}$(\GeV)&$N_s$&$\sigma$&$\varepsilon_{\rm eff}$(\%)\\
\hline\hline
$M_{\pi\pi}<0.39$&$-2.7^{+3.9}_{-3.0}$&-&11.2\\
$0.39-0.46$&$9.5^{+5.9}_{-5.0}$&2.1&11.5\\
$0.46 - 0.53$&$K_S^0$ veto&-&-\\
$0.53-0.6$&$-0.1^{+3.9}_{-3.1}$&-&11.3\\
$0.6-0.67$&$1.9^{+4.9}_{-4.4}$&0.5&11.9\\
$0.67-0.74$&$10.8^{+6.7}_{-5.8}$&2.0&12.1\\
$0.74-0.81$&$13.0^{+6.5}_{-5.6}$&2.6&12.3\\
$0.81-0.88$&$13.9^{+6.1}_{-5.3}$&3.1&11.8\\
$0.88-0.95$&$16.5^{+6.0}_{-5.3}$&4.1&10.8\\
$0.95-1.02$&$0.5^{+2.6}_{-2.1}$&-&9.6\\
$1.02-1.09$&$3.6^{+4.0}_{-3.1}$&1.2&8.4\\
$1.09-1.16$&$1.2^{+3.2}_{-2.8}$&0.5&6.5\\
$1.16-1.22$&$2.3^{+2.9}_{-1.9}$&1.3&3.5\\
\hline\hline
\end{tabular}
\end{center}
\end{table}
\begin{table}[htbp]
\begin{center}
\caption{Yields, statistical significance and efficiencies ($\varepsilon_{\rm eff}$) in different $M_{\pi\pi}$ bin for \PPPIPIC.}
\label{table:impipippr2}
\begin{tabular}[t]{lccc}
\hline\hline
$M_{\pi\pi}$(\GeV)&$N_s$&$\sigma$&$\varepsilon_{\rm eff}$(\%)\\
\hline\hline
$M_{\pi\pi}<0.39$&$-0.5^{+5.3}_{-4.4}$&-&4.3\\
$0.39-0.46$&$3.0^{+8.8}_{-7.8}$&0.3&4.1\\
$0.46-0.53$&$7.5^{+10.0}_{-9.0}$&0.8&4.9\\
$0.53-0.6$&$23.2^{+12.8}_{-11.9}$&2.2&4.7\\
$0.6-0.67$&$-5.9^{+10.5}_{-9.2}$&-&4.8\\
$0.67-0.74$&$25.7^{+12.3}_{-11.4}$&1.8&5.0\\
$0.74-0.81$&$53.9^{+16.5}_{-15.7}$&3.7&5.1\\
$0.81-0.88$&$5.3^{+13.3}_{-12.0}$&0.4&4.8\\
$0.88-0.95$&$-3.0^{+9.8}_{-8.5}$&-&4.3\\
$0.95-1.02$&$20.9^{+11.3}_{-9.8}$&1.7&3.7\\
$1.02-1.09$&$5.8^{+8.1}_{-7.6}$&0.8&2.7\\
$1.09-1.16$&$25.4^{+9.5}_{-8.7}$&3.1&2.7\\
$1.16-1.23$&$6.2^{+7.5}_{-8.4}$&0.8&2.2\\
$1.23-1.3$&$-0.3^{+5.3}_{-4.5}$&-&0.8\\
\hline\hline
\end{tabular}
\end{center}
\end{table}
We attempted to find the contribution of \PPRHC~by minimizing the $\chi^2$ between the observed data
and the assumed non-resonant \PPPIPIC~and \PPRHC~decays. To describe the $M_{\pi\pi}$ distribution, we
use the phase space model for non-resonant \PPPIPIC~ and a Breit-Wigner function convolved
with a Gaussian function for \PPRHC. We set the Breit-Wigner function with
its mean and width to the nominal values for the $\rho^+$ convolved with a
Gaussian resolution function of 5 $\rm$\MeV~width. The result is shown in Fig.~\ref{fig:ppr2chi}.

The fit gives a yield of $86\pm41$ events with a $\frac{\chi^2}{n_{\rm dof}}$ of 17.0/11 for \PPRHC.
Our current data sample is not large enough to separate the contributions of \PPRHC~and
non-resonant \PPPIPIC. The measured $\mathcal{B}$(\PPPIPIC) with \PPRHC~included
is almost a factor of ten smaller than the predicted $\mathcal{B}$(\PPRHC)~\cite{REF29}.\\
\indent There are modes sharing the same final state-particles as our signal, such as
$B\rightarrow\overline{p}\Delta^{++}\pi$ or $B\rightarrow\overline{p}\Lambda^0\pi$.
Examining the $M_{\Delta(p\pi^+)}$ and $M_{\Lambda(p\pi^-)}$ spectra, we find no obvious contribution from
these modes.\\
\indent We investigate the $M_{p\overline{p}}$ distribution of $B$ signals in three
regions:
$M_{p\overline{p}}<2.85$ \GeV~for the threshold enhancement region;
$2.85<M_{p\overline{p}}<3.128$ \GeV~for the charmonium enhanced region; and
3.128 \GeV~$<M_{p\overline{p}}$ for the phase space dominant region.
We perform a 2D ($\Delta E,M_{\rm bc}$) likelihood fit to extract the signal
yields of the $B \rightarrow p \overline{p}\pi\pi$ decays in each region.

\begin{table}[htbp]
\begin{center}
\caption{Yields, statistical significance and efficiencies ($\varepsilon_{\rm eff}$) in different $M_{p\overline{p}}$ bins for \PPPIPI~($0.6<M_{\pi\pi}<$ 1.22 \GeV)}
\label{table:impppprho}
\begin{tabular}[t]{lccc}
\hline\hline
$M_{p\overline{p}}$(\GeV)&$N_s$&$\sigma$&$\varepsilon_{\rm eff}$(\%)\\
\hline\hline
$M_{p\overline{p}}<2.85$&$26.1^{+10.0}_{-9.1}$&4.0&9.8\\
$2.85<M_{p\overline{p}}<3.128$&$19.6^{+10.2}_{-9.3}$&2.9&9.9\\
$3.128<M_{p\overline{p}}$&$29.1^{+16.2}_{-13.1}$&3.5&9.4\\
\hline\hline
\end{tabular}
\end{center}
\end{table}
\begin{table}[htbp]
\begin{center}
\caption{Yields, statistical significance and efficiencies ($\varepsilon_{\rm eff}$) in different $M_{p\overline{p}}$ bins for \PPPIPIC~($M_{\pi\pi}<$ 1.3 \GeV)}
\label{table:imppppr2}
\begin{tabular}[t]{lccc}
\hline\hline
$M_{p\overline{p}}$(\GeV)&$N_s$&$\sigma$&$\varepsilon_{\rm eff}$(\%)\\
\hline\hline
$M_{p\overline{p}}<2.85$&$133.5^{+26.6}_{-25.2}$&5.1&4.8\\
$2.85<M_{p\overline{p}}<3.128$&$12.3^{+10.3}_{-9.7}$&1.4&4.0\\
$3.128<M_{p\overline{p}}$&$-3.8^{+15.1}_{-13.8}$&-&3.4\\
\hline\hline
\end{tabular}
\end{center}
\end{table}

Tables~\ref{table:impppprho} and~\ref{table:imppppr2} show the fitted yields with statistical fit significances
for $B^0 \rightarrow p \overline{p}\pi^+\pi^-$ and $B^+ \rightarrow p \overline{p}\pi^+\pi^0$, respectively.
The charmonium-enhanced region, $2.85<M_{p\overline{p}}<3.128$ \GeV,
includes other expected resonant modes such as $B \rightarrow J/\psi \rho$~\cite{pdg18}.
We find \PPPIPI~events are equally distributed in the bins below and
above the charmonium-enhanced region, while \PPPIPIC~events are dominant in the bin below the charmonium enhanced region.

Sources of systematic uncertainties are summarized in Table~\ref{table:eff}.
The number of $B\overline{B}$ pairs is known to 1.4\%.
By using the partially reconstructed $D^{*+}\rightarrow D^0 \pi^{+}$
with $D^0 \rightarrow \pi^+ \pi^- K^{0}_{S}$ events, the uncertainty due to the
charged-track reconstruction efficiency is estimated to be 0.35\% per track.
We use a $\Lambda \rightarrow p \pi^{-}$ ($D^{*+} \rightarrow D^{0
}\pi^{+}$, $D^{0} \rightarrow K^{-}\pi^{+}$)
sample to calibrate the MC $p$ ($\pi^+$) identification efficiency and assign an uncertainty of 3.3\% and 2.4\% for \PPPIPI~and \PPPIPIC~decays, respectively.
For $\pi^0$ reconstruction, we determine its uncertainty
by using a $\tau^- \rightarrow \pi^-\pi^0\nu$ data sample~\cite{REF25}.
To estimate the systematic error due to continuum suppression,
we use the $B^0 \rightarrow p\overline{p}\overline{D}{}^0$
and $B^0 \rightarrow\overline{D}{}^0 \pi^0$ data/MC samples,
where $\overline{D}{}^0\rightarrow K^+\pi^-$.
We choose the efficiency of the phase space model for \PPPIPI~and the
efficiency of the reweighted phase space model for \PPPIPIC, and estimate
the efficiency uncertainty as a difference of signal efficiencies for
\PPPIPI~in the reweighted phase space model and \PPPIPIC~in the phase space model.
The uncertainty associated with fit parameters is examined by repeating the fit with each parameter varied by
one standard deviation from its nominal value. The resulting difference is taken as the systematic uncertainty.
The assumption of no correlation between $\Delta E$ and $M_{\rm bc}$ is examined by replacing PDF of $B$ signal events with the corresponding 2-D histogram function.
\begin{table}[htbp]
\begin{center}
\renewcommand\arraystretch{1.5}

\caption{Table of systematic uncertainties (\%) for \PPPIPI~and \PPPIPIC.}
\label{table:eff}
\begin{tabular}[t]{ccc}
\hline\hline
Uncertainties&\PPPIPI&\PPPIPIC\\
\hline\hline
$N_{B\overline{B}}$&1.4&1.4\\
Tracking&1.4&1.1\\
$p$/$\pi$ identification&3.3&2.4\\
$\pi^0$ reconstruction&-&2.8\\
Continuum suppression&4.7&4.3\\
Decay model&14.3&8.6\\
$\Delta E$, $M_{\rm bc}$ shape&12.4&10.4\\
\hline
Summary&19.9&14.6\\
\hline\hline
\end{tabular}
\end{center}
\end{table}

In summary, we report the observations of \PPPIPI~and \PPPIPIC~with branching
fractions of $(0.83\pm0.17\pm0.17)\times 10^{-6}$ and $(4.58\pm1.17\pm0.67)\times 10^{-6}$
for $M_{\pi^+\pi^-}<1.22$ \GeV~and $M_{\pi^+\pi^0}<1.3$ \GeV, respectively.
In contrast to the theoretical prediction~\cite{REF29}, the measured $\mathcal{B}$ for \PPPIPIC~in the $\rho$-enhanced region is an order of magnitude smaller than the theoretical expectation.
We find the \PPPIPIC~decay dominated by the lower $M_{p\overline{p}}$ bin, which is not
the case in the \PPPIPI~decay. These findings are useful for the future theoretical investigation.

We thank the KEKB group for the excellent operation of the
accelerator; the KEK cryogenics group for the efficient
operation of the solenoid; and the KEK computer group, and the Pacific Northwest National
Laboratory (PNNL) Environmental Molecular Sciences Laboratory (EMSL)
computing group for strong computing support; and the National
Institute of Informatics, and Science Information NETwork 5 (SINET5) for
valuable network support.  We acknowledge support from
the Ministry of Education, Culture, Sports, Science, and
Technology (MEXT) of Japan, the Japan Society for the
Promotion of Science (JSPS), and the Tau-Lepton Physics
Research Center of Nagoya University;
the Australian Research Council including grants
DP180102629, 
DP170102389, 
DP170102204, 
DP150103061, 
FT130100303; 
Austrian Science Fund (FWF);
the National Natural Science Foundation of China under Contracts
No.~11435013,  
No.~11475187,  
No.~11521505,  
No.~11575017,  
No.~11675166,  
No.~11705209;  
Key Research Program of Frontier Sciences, Chinese Academy of Sciences (CAS), Grant No.~QYZDJ-SSW-SLH011; 
the  CAS Center for Excellence in Particle Physics (CCEPP); 
the Shanghai Pujiang Program under Grant No.~18PJ1401000;  
the Ministry of Education, Youth and Sports of the Czech
Republic under Contract No.~LTT17020;
the Carl Zeiss Foundation, the Deutsche Forschungsgemeinschaft, the
Excellence Cluster Universe, and the VolkswagenStiftung;
the Department of Science and Technology of India;
the Istituto Nazionale di Fisica Nucleare of Italy;
National Research Foundation (NRF) of Korea Grants
No.~2015H1A2A1033649, No.~2016R1D1A1B01010135, No.~2016K1A3A7A09005
603, No.~2016R1D1A1B02012900, No.~2018R1A2B3003 643,
No.~2018R1A6A1A06024970, No.~2018R1D1 A1B07047294; Radiation Science Research Institute, Foreign Large-size Research Facility Application Supporting project, the Global Science Experimental Data Hub Center of the Korea Institute of Science and Technology Information and KREONET/GLORIAD;
the Polish Ministry of Science and Higher Education and
the National Science Center;
the Grant of the Russian Federation Government, Agreement No.~14.W03.31.0026; 
the Slovenian Research Agency;
Ikerbasque, Basque Foundation for Science, Spain;
the Swiss National Science Foundation;
the Ministry of Education and the Ministry of Science and Technology of Taiwan;
and the United States Department of Energy and the National Science Foundation.

\bibliography{refpage3}
\end{document}